\documentclass[preprint,12pt,prb,noshowpacs,a4paper,noshowkeys]{revtex4}
\usepackage{amssymb}
\usepackage{amsmath}
\usepackage{graphicx,float}

\newtheorem{theorem}{Theorem}
\newtheorem{acknowledgement}[theorem]{Acknowledgement}

\begin{document}

\title{Enhanced thermoelectric performance in thin films of
three-dimensional topological insulators}
\author{T. H. Wang and H. T. Jeng}
\email{jeng@phys.nthu.edu.tw}
\affiliation{Department of Physics, National Tsing Hua University, 101 Section 2 Kuang Fu
Road, Hsinchu,Taiwan 30013,R.O.C.}

\renewcommand{\figurename}{ \textbf{FIG.}}

\begin{abstract}
\qquad Thermoelectric (TE) devices have been attracting increasing attention
because of their ability to convert heat directly to electricity. To date,
improving the TE figure of merit remains the key challenge. The advent of
the topological insulator and the emerging nanotechnology open a new way to
design high-performance TE devices. In this paper, we investigate the TE
transport properties of the Bi$_{2}$Se$_{3}$ thin film by the
first-principle calculations and the Boltzmann transport theory. By
comparing our calculations with the earlier experimental data, we
demonstrate that, for the Bi$_{2}$Se$_{3}$ film of thickness larger than six
quintuple layers, the relaxation time of the topological surface states in
the bulk gap is about hundreds of femtoseconds, which is about two orders
larger than that of the bulk states. Such a large relaxation-time difference
causes the ratio of the electrical conductance to the thermal conductance
much larger than the value predicted by the Wiedemann-Franz law, as well as
the large magnitude of Seebeck coefficient, and consequently the large TE
figure of merit, when the Fermi level is near the conduction band edge. We
shows that the TE performance can be further improved by introducing defects
in the middle layers of the thin film. The improvement is generally
significant at room temperature and can be further enhanced at higher
temperature.
\end{abstract}

\date{\today }
\maketitle
\section{Introduction}

 \qquad Today, people faces numerous problems relating to energy supply and
consumption. Nearly $70\%$ of the world energy are wasted and dissipated
into the environments as low grade heat\cite{Zevenhoven(2011)}. Therefore,
the thermoelectric (TE) generators, which can convert heat into electricity,
having advantages of solid-state operation without moving parts, no release
of greenhouse gases, good stability, and high reliability have attracted
widespread research interest \cite%
{Rowe(1955),Tritt(2006),Snyder(2008),Tritt(2011),Alam(2013),He(2015),Aswal(2016),Zhang(2016)}
However, the TE devices available to date are still in the limited
application mainly because of its low energy-conversion efficiency.\bigskip\ 

Topological insulators (TIs), recently attracting great interest, are
materials with an inverted bulk gap induced by the strong spin-orbit
coupling and the metallic topological surface states (TSSs), which are
protected by time reversal symmetry\cite%
{Kane(2005),Bernevig(2006)prl,Fu(2007),Moore(2007),Qi(2008),Hasan(2010),Bansil(2016)}%
. Most TIs such as Bi$_{2}$Te$_{3}$, Sb$_{2}$Te$_{3}$, and Bi$_{2}$Te$_{2}$%
Se are also good TE materials because the TIs and the TE materials generally
share two common features, which are not directly related to the TSSs\cite%
{Zhang(2009),Chen(2009),Xie(2010),Muchler(2013),Shi(2015)}. 1. The corrugated
constant-energy surfaces and the complex band structures, generally can be
caused by the band inversion in TIs, will lead to a high power factor in TE
materials. 2. The large atomic mass, causing the strong spin-orbit coupling
generally necessary to TIs, will bring about a low lattice thermal
conductivity, which is beneficial to the TE performance. When the size is
reduced to the nanometer scale, the TSSs will significantly change the TE
performance for both the two-dimensional (2D) and the three-dimensional (3D)
TIs \cite{ZhangSC(2014),Liang(2016)}. Layered bismuth selenide (Bi$_{2}$Se$%
_{3}$) is one of the most intensively studied 3D TIs. It has a considerable
bulk gap ($\sim 0.3$ eV) and topologically protected metallic surface
states\cite{Zhang(2009),Xia(2009),Hsieh(2009)}. Although the bulk
crystalline Bi$_{2}$Se$_{3}$ is not a good TE material, it has been shown
that the TE performance can be greatly enhanced by fabricating a
single-layer-based composite made from atomically thick layers\cite%
{Sun(2012)}.

 In this paper, we systematically study the TE performance of the Bi$_{2}$Se$%
_{3}$ thin film through the first-principle calculations and the Boltzmann
transport theory. We mainly focus on the case of the film thickness larger
than six quintuple layers (QLs) so that, as discussed later, the relaxation
time of the TSSs in the bulk gap is about two orders of magnitude larger
than that of the bulk states. Therefore, we will call the TSSs in the bulk
gap the long lifetime states (LLSs) and the other states the short lifetime
states (SLSs). We will explicate the effect of the large relaxation ratio
between the LLSs and the SLSs on the TE parameters, and how it improves the
TE performance. The TE performance will be further improved when the SLS
relaxation time $\tau _{SLS}$ is shorten without changing the LLS relaxation
time $\tau _{LLS}$. This can be achieved by introducing defects in the
middle layers as discussed later. Unlike the common cases of small-gap
materials, in which the bipolar effect can significantly deteriorates the TE
performance at high temperature, such improvement of the TE performance is
significant at room temperature and further enhanced at higher temperature.
Although we only consider the case of Bi$_{2}$Se$_{3}$ thin film, similar
enhancement of the TE performance is expected in some of the other 3D TI thin
films.

\section{Method}

\subsection{Electronic Structure}

 \qquad The electronic structure is calculated through the projector augmented wave
(PAW) approach within the framework of density functional theory (DFT) as
implemented in the Vienna Ab initio Simulation Package (VASP)\cite%
{Kresse(1995),Kresse(1996)a,Kresse(1996)b}. The exchange-correlation is
described in the Perdew-Burke-Ernzerhof (PBE) form of generalized gradient
approximation (GGA)\cite{Perdew(1992)a,Perdew(1992)b}. The spin-orbit
coupling is taken into account. The $13\times 13\times 1$ Monkhorst--Pack
mesh is used for $k$-point sampling within the Brillouin zone. The cutoff
energy for plane wave basis is set as $450$ eV. The energy convergence
threshold is set to $10^{-9}$ eV in the self-consistent calculation. For
structure relaxation, the van der Waals interactions between two adjacent
quintuple layers are included using the DFT-D3 method with Becke-Jonson
damping\cite{Grimme(2010),Grimme(2011)}. All the internal atomic coordinates
and the lattice constant are relaxed until the magnitude of the force acting
on all atoms is less than $0.5$ meV$/$\text{\AA}.

\subsection{Transport properties}

 \qquad The efficiency of TE devices is determined by the dimensionless figure of
merit\cite{Rowe(1955)}. For the case that the applied field of the thin film
is along the in-plane direction, it can be expressed as

\begin{equation}
zT=\frac{\sigma S^{2}T}{\kappa }  \label{Eq01}
\end{equation}%
where $T$ is the absolute temperature, $S$ the Seebeck coeficiant, $\sigma $
the electrical sheet conductance, and $\kappa $ the thermal sheet
conductance, which is the sum of its electronic part $\kappa _{e}$ and its
lattice part $\kappa _{L}$. Except for the lattice thermal conductance $%
\kappa _{L}$, all the TE parameters can be obtained from the Boltzmann
transport equation with the relaxation time approximation. They can be
expressed as%
\begin{eqnarray}
\sigma  &=&\frac{e^{2}}{\hbar }I_{0},  \label{Eq02a} \\
S &=&-\frac{k_{B}}{e}\frac{I_{1}}{I_{0}},  \label{Eq02b} \\
\kappa _{e} &=&\frac{k_{B}^{2}T}{\hbar }\left( I_{2}-\frac{I_{1}^{2}}{I_{0}}%
\right) ,  \label{Eq02c}
\end{eqnarray}%
where $e$ is the elementary charge, $k_{B}$ the Boltzmann constant, $\hbar $
the reduced Planck constant. The $I_{n}$ is a dimensionless integral which
can be written as%
\begin{equation}
I_{n}=\int d\varepsilon \left( -\frac{\partial f_{0}}{\partial \varepsilon }%
\right) \left( \varepsilon -\varepsilon _{f}\right) ^{n}\Sigma ,
\label{Eq03}
\end{equation}%
where $f_{0}=1/\left[ \exp \left( \varepsilon -\varepsilon _{f}\right) +1%
\right] $ is the Fermi-Dirac distribution function, $\varepsilon _{f}$ the
reduced Fermi level, and $E_{f}$ the Fermi level. The $\Sigma $ is the
dimensionless transport distribution function (TDF)\cite{Mahan(1996)}, which
can be expressed as%
\begin{equation}
\Sigma =\frac{\hbar }{\left( 2\pi \right) ^{2}}\int d^{2}k\delta \left( E-E_{%
\mathbf{k}}\right) v_{\mathbf{k}}^{2}\tau _{\mathbf{k}},  \label{Eq04}
\end{equation}%
where $E_{\mathbf{k}}$ is the electronic band structure obtained from the
DFT calculations, $v_{\mathbf{k}}$ the group velocity in the direction of
the applied field, and $\tau _{\mathbf{k}}$ the relaxation time.

The lattice thermal conductance $\kappa _{L}$ can be obtained through
Callaway's model. It can be expressed as\cite{Callaway(1959),Navratil(2004)}%
\begin{equation}
\kappa _{L}=\frac{k_{B}d}{2\pi ^{2}v_{s}}\left( \frac{k_{B}T}{\hbar }\right)
^{3}\int_{0}^{T_{D}/\sqrt[3]{5}T}dy\tau _{c}\frac{y^{4}e^{y}}{\left(
e^{y}-1\right) ^{2}},  \label{Eq05}
\end{equation}%
where $d$ is the film thickness, $v_{s}$ the sound velocity, $T_{D}$ the
Debye temperature, $y$ proportional to the phonon frequency $\omega $ the
dimensionless parameter $y=\hbar \omega /k_{B}T$. The $\tau _{c}$ is the
phonon scattering relaxation time, which can be written as%
\begin{equation}
\tau _{c}^{-1}=\frac{v_{s}}{d}+A\omega ^{4}+B\omega ^{2}T\exp \left( -\frac{%
\Theta _{D}}{CT}\right) ,  \label{Eq06}
\end{equation}%
where $A=9.42\times 10^{-42}$ (s$^{3}$), $B=7.78\times 10^{-18}$ (sK$^{-1}$)
and $C=2.8$ are parameters independent of temperature\cite{Navratil(2004)}. The three terms of
equation (\ref{Eq06}) represent, respectively, the boundary scattering, the
point defect (Rayleigh) scattering, and the phonon-phonon scattering.

\section{Results and Discussion}

\subsection{Electronic Structure \& Electron Scattering}

 \qquad The electronic structure of the Bi$_{2}$Se$_{3}$ thin film of thickness $8$
QLs is shown in Fig. 1b, in which we can see the massless Dirac cone with
the Dirac point at zero energy. The Dirac-cone states are doubly degenerate
states, which can be regarded as two TSSs on different sides of the film. As
can be seen, there is a bulk gap, the energy region between the valence band
edge (VBE) $E_{v}=-0.5$ meV and the conduction band edge (CBE) $E_{c}=295$ meV
indicated by the yellow shaded region, in which there is no electronic
states except for the TSSs. The wave functions of the TSSs on different
sides of the film are spatially separated so that the orbital contribution
from the atoms in the middle $2$ QLs (indicated by the red box in Fig. 1a)
is less than $1\%$. The TSSs in the bulk gap (hereafter call the LLSs)
should have long relaxation times due to two reasons. (i) The large-angle
scattering between the TSSs on the same side of the thin film is suppressed
because the spin polarization of the TSSs circulates around the $\Gamma $
point as can be seen in Fig. 1c. (ii) The scattering between the TSSs on the
opposite sides of the thin film is negligible because of the spatial
separation between the initial and final states as mentioned above. The
relaxation times of the other states (hereafter call the SLSs), $\tau _{SLS}$%
, should be much smaller than that of the LLSs, $\tau _{LLS}$, because the
electrons of the SLSs can be elastically scattered into the bulk states. The
long relaxation time of the TSSs in the bulk gap (i.e., the LLSs) can be
demonstrated by comparing our computed value of the sheet conductance with
the experimental data in ref.~\citenum{Taskin(2012)} (see Supplementary
Information for more details). The estimated value is $230$ fs, which is
about two orders higher than that of bulk relaxation time ($\sim 2.7$ fs)%
\cite{Luo(2012)}. In this paper, we adopt the dual relaxation time model,
which is a generalization of the conventional constant relaxation time
approximation\cite{ZhangSC(2014)}. In this model, the relaxation times $\tau
_{LLS}$ and $\tau _{SLS}$ are two constants. The integral $I_{n}$ is divided
into the LLS part $I_{n}^{LLS}$ and the SLS part $I_{n}^{SLS}$. Then we can
define the LLS (SLS) electrical conductance $\sigma _{LLS}$ ($\sigma _{SLS}$
), and the LLS (SLS) Seebeck coefficient $S_{LLS}$ ($S_{SLS}$ ) by
the equations the same as equations (\ref{Eq02a}) and (\ref{Eq02b}), except that
the integral is replaced by the $I_{n}^{LLS}$ ($I_{n}^{SLS}$). According to
this definition, the electrical conductance is the sum of the LLS and the
SLS conductances\cite{Goldsmid(book)},%
\begin{equation}
\sigma =\sigma _{LLS}+\sigma _{SLS}.  \label{Eq07}
\end{equation}%
On the other hand, the Seebeck coefficient is a weighted average of the
Seebeck coefficients associated with the two type of the carriers,%
\begin{equation}
S=\frac{\sigma _{LLS}S_{LLS}+\sigma _{SLS}S_{SLS}}{\sigma _{LLS}+\sigma
_{SLS}}  \label{Eq08}
\end{equation}

\subsection{Thermoelectric transport properties}

 \qquad The room-temperature TE figure of merit of the $8$QL Bi$_{2}$Se$_{3}$ film
as a function of the Fermi level is shown in Fig. 2a. Here, the relaxation
time of the LLSs and the SLSs are, respectively, set to be $230$ fs and $1$
fs. As can be seen, there are four $zT$ peaks denoted by P1-P4, among which
the P1 (P4) is located away from the VBE (CBE). When the Fermi level is near
the P1 or P4 energy, the conduction carriers are almost the SLS carriers. On
the other hand, the peak P2 (P3) is located near the VBE (CBE). When the
Fermi level is around the P2 (P3) energy, both the LLSs and the SLSs are
significant to the TE performance. The $zT$ peak values of P2 and P3 are
much greater than those of P1 and P4. Hereafter, we will mainly focus on P2
and P3. Sometimes, we may only discuss the case of P3, and the discussion of
P2 is similar to that of P3 unless otherwise mentioned.

Figure 2b shows the electrical sheet conductance as a function of the Fermi
level. As can be seen, when the Fermi level is located in the bulk gap, the
magnitude of the total conductance is nearly equal to that of the LLS
contribution. Because the LLS density of states (DOS) increases with the
Fermi level, the conductance increases with the Fermi level to the maximum
at the energy about $2.5k_{B}T$ (or 65 meV) below the CBE. It then decreases
because of the decline in the conduction carrier number of the LLSs
regardless the increase in that of the SLSs. When the Fermi level is at the
P3 energy, which is $42$ meV above the CBE, the conduction carrier number of
the SLSs is much larger than that of the LLSs. Nevertheless, the total
conductance is mainly contributed from the LLS conductance $\sigma _{LLS}$
because of the large relaxation time difference between the LLSs and the
SLSs.

The Seebeck coefficient as a function of the Fermi level is shown in Fig.
2c. Similar to the typical semiconductor cases, the SLS Seebeck coefficient $%
S_{SLS}$ is negative for $n$-type doping ($E_{f}\gtrsim E_{c}$) and positive
for $p$-type doping ($E_{f}\lesssim E_{v}$ ). When the Fermi level is in the
bulk gap and away from the midgap, the magnitude of the $S_{SLS}$ increases
as the Fermi level moves toward the midgap. This can be easily comprehended
by rewriting the Seebeck coefficients of the LLSs and the SLSs as%
\begin{equation}
S_{i}=-\frac{\left\langle E\right\rangle _{i}-E_{f}}{eT},  \label{Eq09}
\end{equation}%
where%
\begin{equation}
\left\langle E\right\rangle _{i}=\frac{\int_{\Delta _{i}}d\varepsilon \left(
-\frac{\partial f_{0}}{\partial \varepsilon }\right) \Sigma E}{\int_{\Delta
_{i}}d\varepsilon \left( -\frac{\partial f_{0}}{\partial \varepsilon }%
\right) \Sigma }  \label{Eq10}
\end{equation}%
is the TDF-weighted average energy of the conduction carriers, and $\Delta
_{LLS}$ ($\Delta _{SLS}$) the energy region of the LLSs (SLSs), i.e. the
yellow (white) shaded region in Figs. 1b and 2. For $n$-type ($p$-type)
doping, almost all the conduction carriers and hence the $\left\langle
E\right\rangle _{SLS}$ are slight higher (lower) than the CBE (VBE) so that
the magnitude of the $S_{SLS}$ increase as the Fermi level moves toward the
midgap. Due to the bipolar effect, the magnitude of the $S_{SLS}$ reduces
near the middle of the gap\cite{Goldsmid(book)}. The LLS Seebeck coefficient 
$S_{LLS}$ generally has a sign opposite to the SLS Seebeck coefficient $%
S_{SLS}$. The magnitude of the LLS Seebeck coefficient increases as the
Fermi level moves away from the energy region of the LLSs. This can also be
comprehended from equation (\ref{Eq09}) by considering the fact that the $%
\left\langle E\right\rangle _{LLS}$ is located in the energy region $\Delta
_{LLS}$. The energy of $S_{LLS}=0$ is higher than the midgap energy because
the LLS DOS increases with energy. As defined in equation (\ref{Eq08}), the
Seebeck coefficient is a conductance-weighted average of the LLS and the SLS
Seebeck coefficients. In the limit $\tau _{SLS}\ll \tau _{LLS}$, the Seebeck
coefficient is nearly equal to LLS Seebeck coefficients, whose magnitude is
extraordinarily large when the Fermi level is away from the energy region $%
\Delta _{LLS}$. In the present case, the relaxation time of the SLSs is
indeed much smaller than that of the LLSs. The resultant magnitude of the
Seebeck coefficient at P3 (P2) is nearly the same as magnitude of the $%
S_{LLS}$ and it is as large as $0.21$ ($0.16$) mV/K.

The ratio of the electrical conductance to the electronic thermal
conductance $\sigma /\kappa _{e}$ as a function of the Fermi level is shown
in Fig. 2d. It is normalized in unit of $R_{WF}=3e^{2}/\pi ^{2}k_{B}^{2}T$,
which is the value given by the Wiedemann-Franz law\cite{Mermin(book)}. As
can be seen, the Wiedemann-Franz law is generally applicable (i.e., $\sigma
/\kappa _{e}\sim R_{WF}$) except for the energy around the band edges. The
Wiedemann-Franz law can be derived by applying the Sommerfeld expansion on
equation (\ref{Eq03}) and retaining the lowest non-vanishing order\cite%
{Mermin(book)}. Because the relaxation time of the electrons near the band
edge and the corresponding TDF vary rapidly with the energy, the
higher-order terms are significant. By taking into account the higher-order
terms and replacing $\Sigma $ with $\int d\varepsilon \left( -\frac{\partial
f_{0}}{\partial \varepsilon }\right) \Sigma $, the ratio $\sigma /\kappa
_{e} $ can be approximated as%
\begin{equation}
\frac{\sigma }{\kappa _{e}}\approx R_{WF}\left[ 1+\frac{\pi ^{2}}{3}\left( 
\frac{k_{B}T\sigma ^{\prime }}{\sigma }\right) ^{2}-\frac{8\pi ^{2}}{15}%
\frac{\left( k_{B}T\right) ^{2}\sigma ^{\prime \prime }}{\sigma }\right] ,
\label{Eq11}
\end{equation}%
where the prime denotes the derivative with respect to $E_{f}$. The first
term is just the result of the Wiedemann-Franz law.
In the energy region slightly below the CBE, the electrical conductance decreases rapidly with energy.
The second term becomes significant, and the ratio $\sigma /\kappa _{e}$ increases with energy to a value much larger than $R_{WF}$.
When the energy further increases, the
dominant type of the conduction carriers changes from the LLS to the SLS and
the sign of the $\sigma ^{\prime }$ changes from negative to positive. This
leads to the decrease of the magnitude of the $\sigma ^{\prime }$ and a
considerable positive $\sigma ^{\prime \prime }$. As a result, the ratio $%
\sigma /\kappa _{e}$ decreases with energy to a value smaller than $R_{WF}$.
Because the magnitude of the Seebeck coefficient increases with energy, the
peak P3 is always located at the energy region where the ratio $\sigma
/\kappa$ decreases with energy. The ratio $\sigma /\kappa _{e}$ at P3
energy is more than twice larger than the value given by the Wiedemann-Franz
law. Unlike the case of $E_{f}\sim E_{c}$, the ratio $\sigma /\kappa _{e}$ shows no significant
increase as $E_{f}\sim E_{v}$. This is mainly because the energy of the
Dirac point is nearly the same as that of the VBE. The LLS DOS near the VBE
is much smaller than that near the CBE. Therefore, the variation of the TDF
with energy near the VBE is not as significant as that near the CBE. This is
also the reason why the figure of merit at P2 is smaller than that at P3 and
why the P2 is located in the energy region $\Delta _{LLS}$ while the P3 is
in the $\Delta _{SLS}$.

\subsection{Further Enhancement by shortening the SLS relaxation time}

 \qquad The TE performance can be further enhanced by reducing the SLS relaxation
time. This can be done without changing the LLS relaxation time by
introducing defects in the middle layers as mentioned earlier. As an
example, we calculate the TE parameters at temperature $T=600$ K as shown in
Fig. 3. Figure 3a-e shows the results of $\tau _{SLS}=2$ fs and Fig. 3k-o
the results of $\tau _{SLS}=0.2$ fs. The TE parameters as functions of the
SLS relaxation time in the range from $0.2$ to$\ 2$ fs for $E_{f}=0.34$ eV
(the P3 energy for $\tau _{SLS}=2$ fs) are shown in Fig. 3f-j. As can be
seen in Fig. 3b and 3l, when the SLS relaxation time reduces from $2$ fs to $%
0.2$ fs, the SLS conductance becomes negligible around the P3 energy. This
causes the increase (decrease) of the magnitude of $\sigma ^{\prime }$ ($%
\sigma ^{\prime \prime }$). The increase (decrease) of the magnitude of $%
\sigma ^{\prime }$ ($\sigma ^{\prime \prime }$) leads to the increase of the
ratio $\sigma /\kappa _{e}$ as shown in Fig. 3i. Furthermore, the magnitude
of the Seebeck coefficient increases with the decreasing SLS relaxation
time. It is nearly equal to the magnitude of LLS Seebeck coefficient when $%
\tau _{SLS}=0.2$ fs as shown in Fig. 3j. These cause the remarkable increase
in the figure of merit at P3 energy from $1.31$ to $2.94$ (see Fig. 3f) in
spite of the slight decrease of the electrical conductance (see Fig. 3g).
The reduction of the SLS relaxation time results in the blue shift of the P3
energy as indicated by the blue and green dashed lines in Figs 3a, k. This
further enhances the figure of merit to the value of $zT=3.8$.
In addition, we perform similar calculation for the room temperature (see Fig. S3). We find the figure of merit is also significantly enhanced.
It increases from $1.67$ to $2.77$ when the SLS relaxation time decreases from $2$ fs to $0.2$ fs.

Figure 4 is a low-temperature ($T=100$ K) counterpart of Fig. 3. As can be
seen, the magnitude of the figure of merit only slightly increases from $%
0.90 $ to $1.06$ when the SLS relaxation time reduces from $2$ fs\ to $0.2$
fs. In addition, it would be difficult to experimentally tune the Fermi
level in such a narrow peak width ($\sim 0.02$ eV) at low temperature as can
be seen in Fig 3a, k. Unlike the high-temperature ($T=600$ K) case, in which
the electronic thermal conductance is generally much larger than the lattice
thermal conductance as shown in Fig. 3c, h and m, the electronic thermal
conductance at low temperature ($T=100$ K) near the P3 energy is much
smaller than the lattice thermal conductance as shown in Fig. 4c, h and m.
Consequently, the ratio $\sigma /\kappa $ is nearly a constant even though
the ratio $\sigma /\kappa _{e}$\ increases significantly with the decrease
of the SLS relaxation time as shown in Fig. 4i.

\subsection{Thickness Dependence of the TE Performance}

 \qquad Figure 5a shows the figure of merit of the $7$, $8$ and $10$QL Bi$_{2}$Se$%
_{3}$ thin films at room temperatue. As can be seen, the figure of merit of
P2 and P3 decreases with the film thickness. Figure 5b is the same as Fig.
5a except that the lattice thermal conductance $\kappa _{L}$ of the $7$QL
and the $9$QL films is set at the value the same as that of the $8$QL film.
We find the figure of merit difference between the films of different
thickness is greatly reduced indicating that the decrease of the figure of
merit with increasing thickness is mainly due to the increase of the lattice
thermal conductance. As we mentioned earlier, the TE performance can be
enhanced by introducing defects into the middle layers of the thin film so
that the SLS relaxation time can be reduced without changing the LLS
relaxation time. For the case of larger film thickness, the low SLS
relaxation time should be more easily to achieve. However, the thermal
conductance increases at the same time. As we discussed earlier, the figure
of merit will barely increase with decreasing SLS relaxation time if the
electronic thermal conductance $\kappa _{e}$ is much smaller than the
lattice thermal conductance $\kappa _{L}$. Therefore, the optimized film
thickness should be the smallest thickness that the condition $\kappa
_{e}\ll \kappa _{L}$ can be satisfied by introducing defects. In addition,
it should not be smaller than $6$ QLs as discussed below.

Thus far, we consider the case of the film thickness larger than $6$ QLs.
When the film thickness is smaller than $6$ QLs, the LLS relaxation time
will be dramatically reduced by the scattering between the TSSs on the
opposite sides as demonstrated in Supplementary Information. In this case,
introducing defects into the Bi$_{2}$Se$_{3}$ thin film will significantly
reduce not only the SLS relaxation time $\tau _{SLS}$ but also the LLS
relaxation time $\tau _{LLS}$. Figure 5c shows the figure of merit of the $5$%
QL Bi$_{2}$Se$_{3}$ thin film for the LLS relaxation times shorter than $230$
fs in the case of the SLS relaxation time $\tau _{SLS}=1$ fs at temperature $%
T=600$. The corresponding $zT$ maximum for $n$-type ($p$-type) doping as a
function of the LLS relaxation time is shown in Fig. 5d. Here, the $n$-type (%
$p$-type) doping is defined as the maximum $zT$ in the case of the Fermi
level higher (lower) than $0.2$ eV. As can be seen, the figure of merit
generally decreases with decreasing LLS relaxation time. For $n$-type
doping, the $zT$ maximum reduces from $2.9$ to $0.1$ when the LLS relaxation
time decreases from $230$ fs to $12$ fs. When the LLS relaxation is reduced
to about $15$ fs, the LLS conductance is comparable with the SLS
conductance causing a strong cancelation between the LLS Seebeck coefficient
and the SLS Seebeck coefficient as indicated by equation (\ref{Eq08}). The
relatively small relaxation time ratio $\tau _{LLS}/\tau _{SLS}$ also
deteriorates the TE performance by reducing the ratio of $\sigma /\kappa _{e}
$ as indicated in equation (\ref{Eq11}). When the LLS relaxation time
further decreases, the figure of merit at P4 (P1) is larger than that at P3
(P2). In this case, the $zT$ maximum increases with decreasing $\tau _{LLS}$
because the bulk carriers dominate the transport when the Fermi level is at
P4 (P1) energy. Finally, when the LLS relaxation time is reduced to $1$ fs
(i.e., $\tau _{LLS}=\tau _{SLS}$), the P2 and P3 disappear, and only the P1
and P4 remain. The quite large $zT$ maximum for $p$-type doping is mainly
due to the large DOS of the valence bands.

\section{Conclusions}

 \qquad In summary, we investigate the TE properties of the Bi$_{2}$Se$_{3}$ thin
film through the first-principle calculations and the Boltzmann transport
theory. When the film thinness is so large ($>6$ QLs) that the coupling
between the TSSs on the opposite sides of the film is negligible, the
large-angle scattering is strongly suppressed, and the relaxation time of
the LLSs can be hundreds of times larger than that of the SLSs as
demonstrated by comparing our calculations with the earlier experimental
data (see Supplementary Information). As a result, the reduction of the
magnitude of the Seebeck coefficient near the P3 (P2) energy caused by the
SLS carriers is negligible and the Seebeck coefficient is nearly equal to
the LLS Seebeck coefficient. The ratio of the electrical conductance to the
electronic thermal conductance can be dramatically deviated from the value
given by the Wiedemann-Franz law $R_{WF}$ when the Fermi level is near the
band edges and much larger than $R_{WF}$ when the Fermi level is at the P3
energy. The TE performance can be further enhanced by reducing the SLS
relaxation time without changing LLS relaxation time. This enhancement is
prominent especially at high temperature and achievable by introducing
defects in the middle layers. When the film thickness increases to $10$ QLs,
the figure of merit decreases mainly due to the increase of the thermal
conductance. When the film thickness decreases to the thickness smaller than 
$6$ QLs, the TE performance will be greatly deteriorated because of the
effective coupling between the TSSs on the opposite sides of the thin film.
The optimized film thickness would be the smallest one, that can satisfy the
condition $\kappa _{e}\ll \kappa _{L}$ by introducing defect in the region more than $3$ QLs away from the surface,
and should be not smaller than $6$ QLs.
Although we only consider the thin films made of Bi$_{2}$Se$_{3}$, the high TE
performance caused by the coexistence of the LLS and SLS conduction carriers
is expected in some of the other TI thin films.

\begin{acknowledgement}
This work was supported by the Ministry of Science and Technology, Taiwan.

\end{acknowledgement}

\bigskip

\begin{figure}[H]
\includegraphics [width=16.5cm]{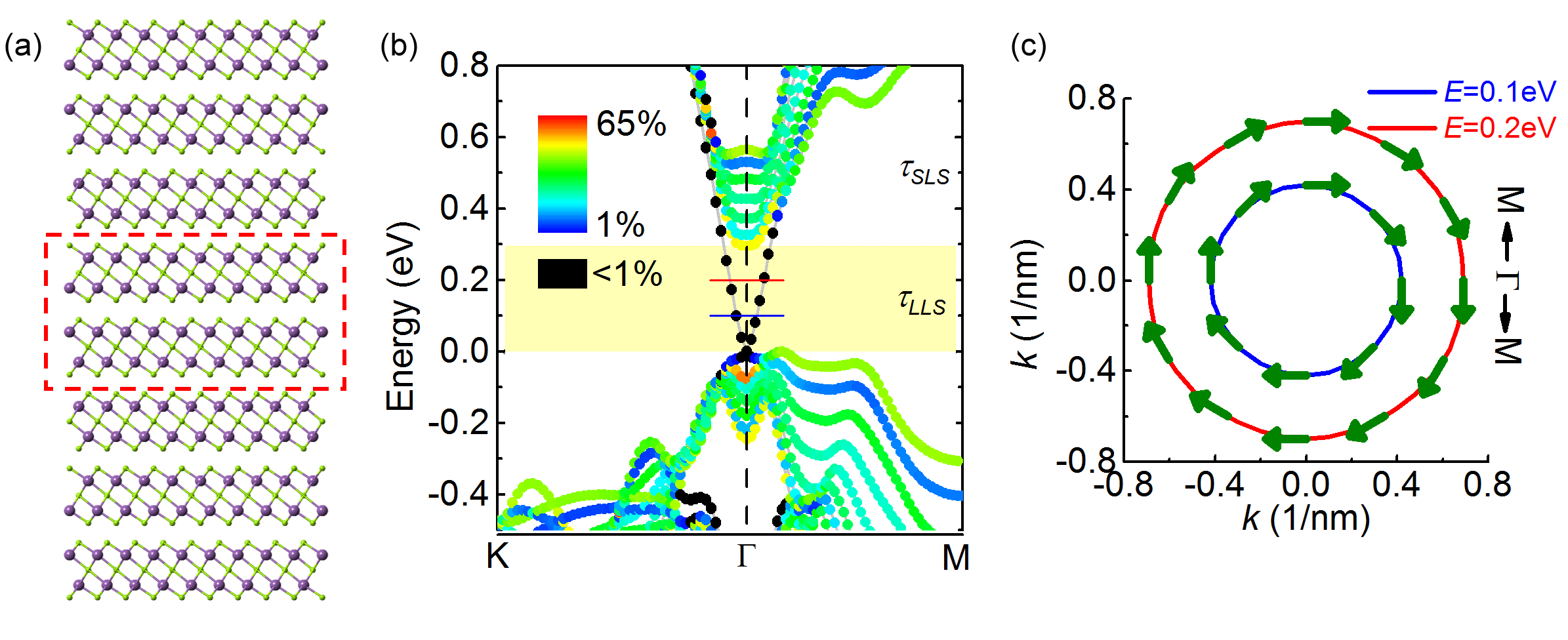}
\caption{\textbf{The crystal structure and the electronic band structure of the }%
$8$\textbf{QL Bi}$_{2}$\textbf{Se}$_{3}$\textbf{. }(\textbf{a}) The crystal
structure. (\textbf{b}) The electronic band structure. The yellow shaded
region denotes of energy region of the LLSs, which is also the bulk gap. The
color of the data points represents the orbital contribution from the atoms
in the middle $2$ QLs indicated by the red box in (\textbf{a}). When it is
black, the orbital contribution from the atoms in the middle $2$ QLs is less
than $1\%$. (\textbf{c}) The spin texture of the constant energy contours.}%
\end{figure}

\newpage

\begin{figure}[H]
\includegraphics [width=14cm]{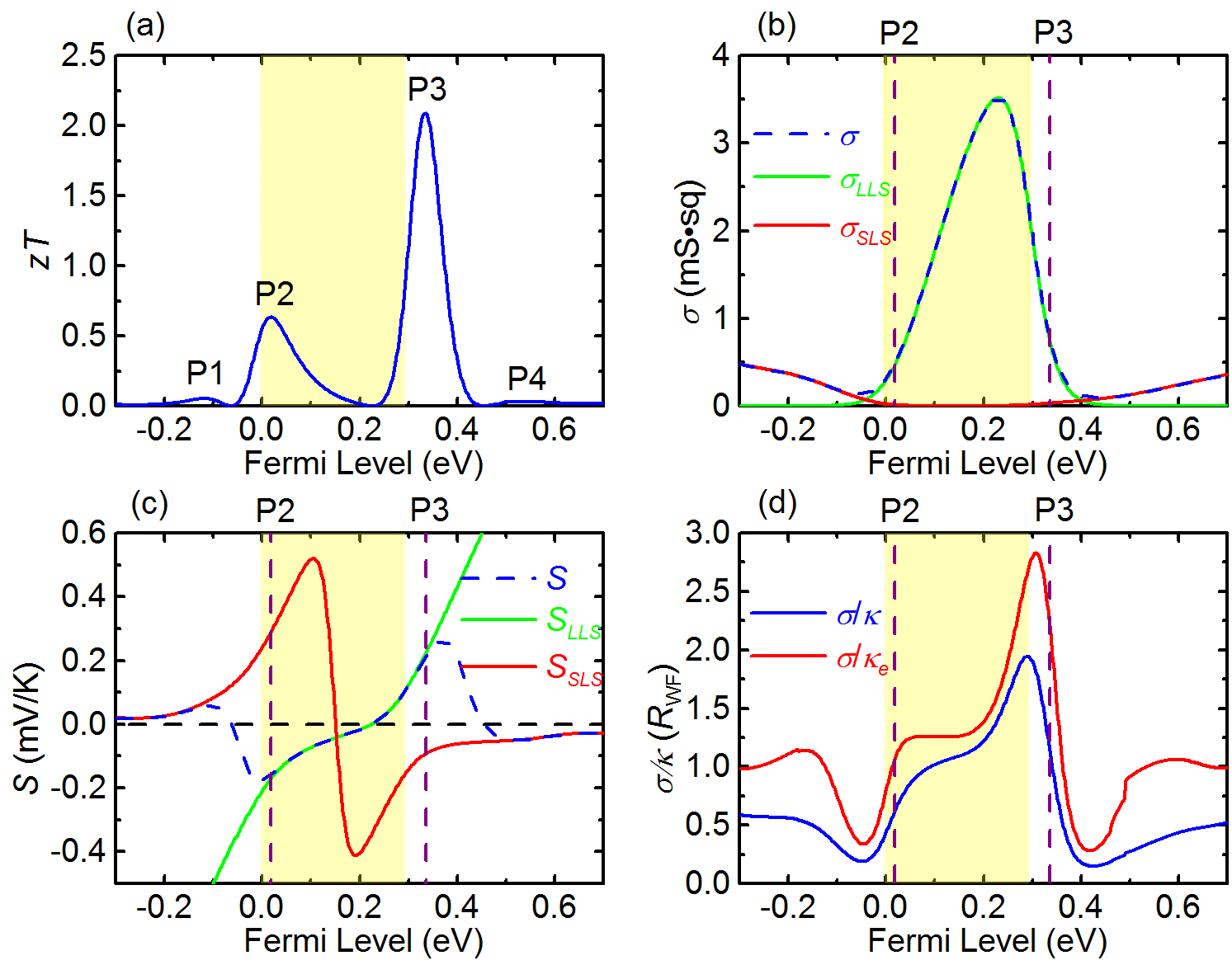}
\caption{\textbf{The room-temperature thermoelectric parameters of the }$8$%
\textbf{QL Bi}$_{2}$\textbf{Se}$_{3}$\textbf{\ as functions of the Fermi
level.} (\textbf{a}) The figure of merit. (\textbf{b}) The electrical sheet
conductance $\sigma $ and the LLS (SLS) conductance $\sigma _{LLS}$ ($\sigma
_{SLS}$). (\textbf{c}) The Seebeck coeffcient $S$ and the LLS (SLS) Seebeck
coeffcient $S_{LLS}$ ($S_{SLS}$). (\textbf{d}) The ratio of $\sigma $ to the
(electronic) thermal sheet conductance $\kappa $ ($\kappa _{e}$). The
relaxation time of the LLSs and the SLSs are, respectively, set to be $230$
fs and $1$ fs. The vertical dashed lines in (\textbf{b}), (\textbf{c}), and (%
\textbf{d}) indicate the P2 and P3 energies.}
\end{figure}

\begin{figure}[H]
\includegraphics [width=17.78cm]{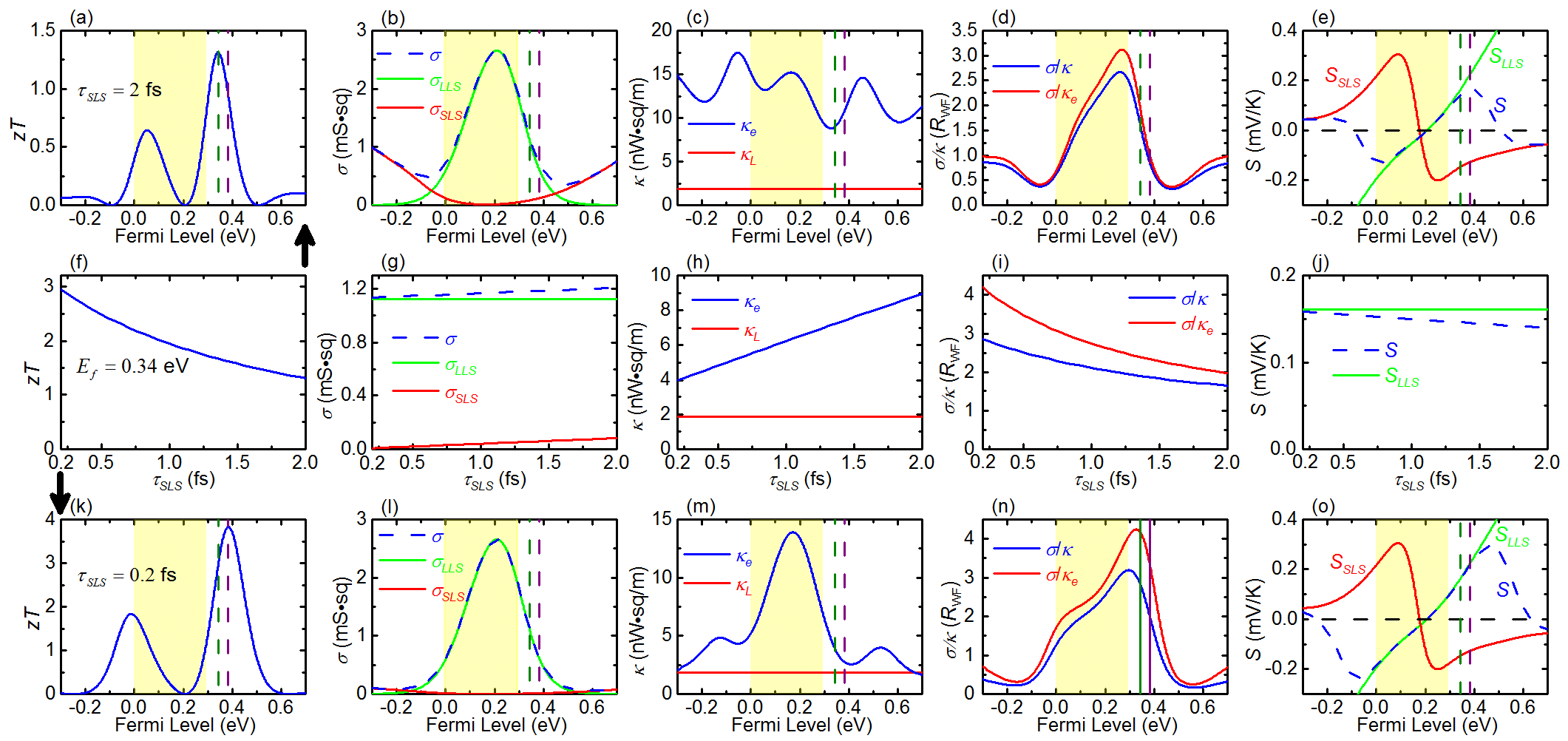}
\caption{\textbf{The thermoelectric parameters of the }$8$\textbf{QL Bi}$_{2}$%
\textbf{Se}$_{3}$\textbf{\ at temperature }$T=600$\textbf{\ K.} (\textbf{a-e}%
) The thermoelectric parameters as functions of the Fermi level for the SLS
relaxation time being set at $\tau _{SLS}=2$ fs and (\textbf{k-o}) those for 
$\tau _{SLS}=0.2$ fs. In (\textbf{a-e}) and (\textbf{k-o}), the vertical
green line and the vertical purple line indicate, respectively, the P3
energy for $\tau _{SLS}=2$ fs and that for $\tau _{SLS}=0.2$ fs. (\textbf{f-j%
}) The thermoelectric parameters as functions of SLS relaxation time when
the Fermi level is set at $0.34$ eV, the P3 energy for $\tau _{SLS}=2$ fs.}%
\end{figure}

\begin{figure}[H]
\includegraphics [width=17.78cm]{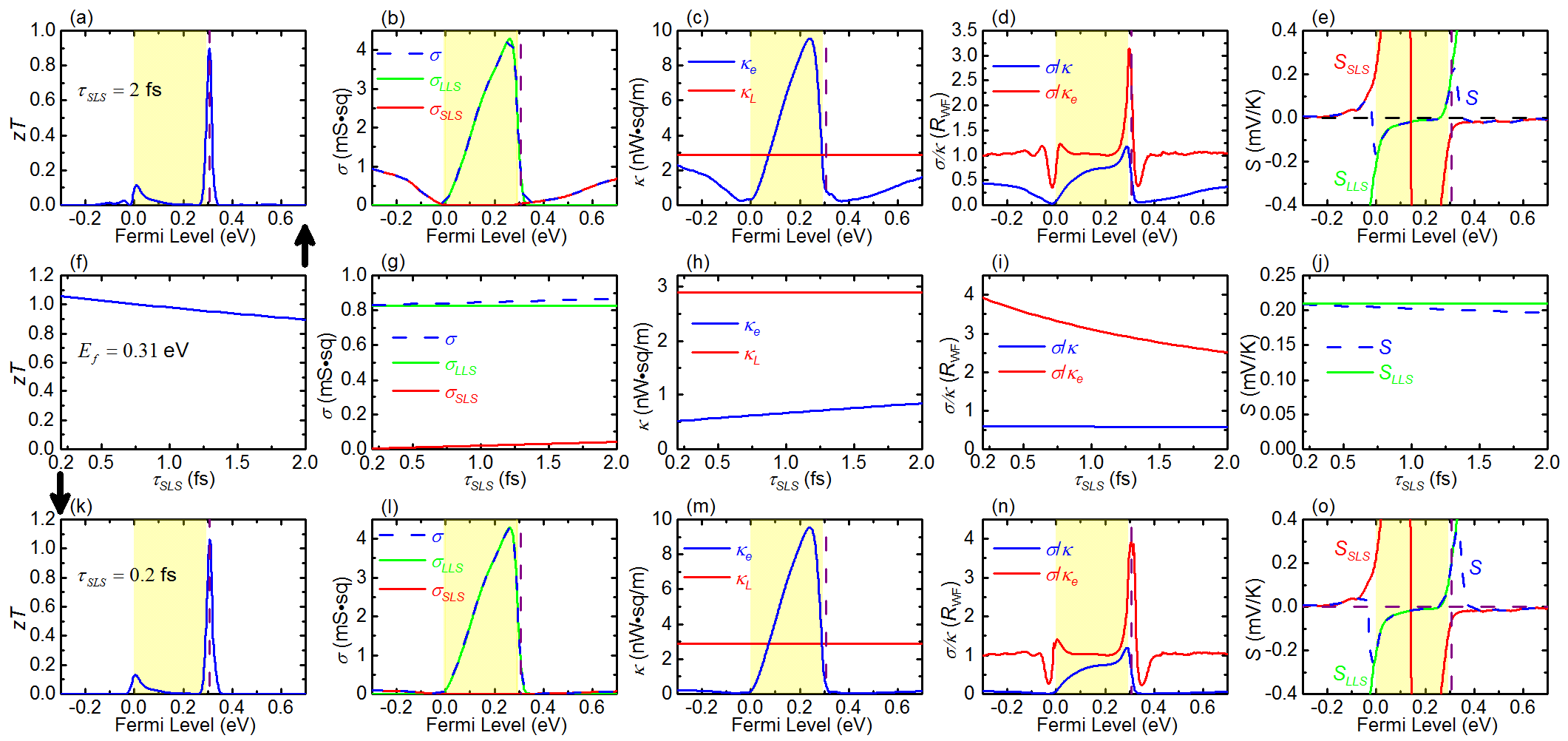}
\caption{\textbf{The thermoelectric parameters of the }$8$\textbf{QL Bi}$_{2}$%
\textbf{Se}$_{3}$\textbf{\ at temperature }$T=100$\textbf{\ K.} (\textbf{a-e}%
) The thermoelectric parameters as functions of the Fermi level for the SLS
relaxation time being set at $\tau _{SLS}=2$ fs and (\textbf{k-o}) those for 
$\tau _{SLS}=0.2$ fs. (\textbf{f-j}) The thermoelectric parameters as
functions of SLS relaxation time when the Fermi level is set at the P3
energy, \ $0.31$ eV, which is denoted in (\textbf{a-e}) and (\textbf{k-o})
by the vertical purple lines.}%
\end{figure}

\begin{figure}[H]
\includegraphics [width=14cm]{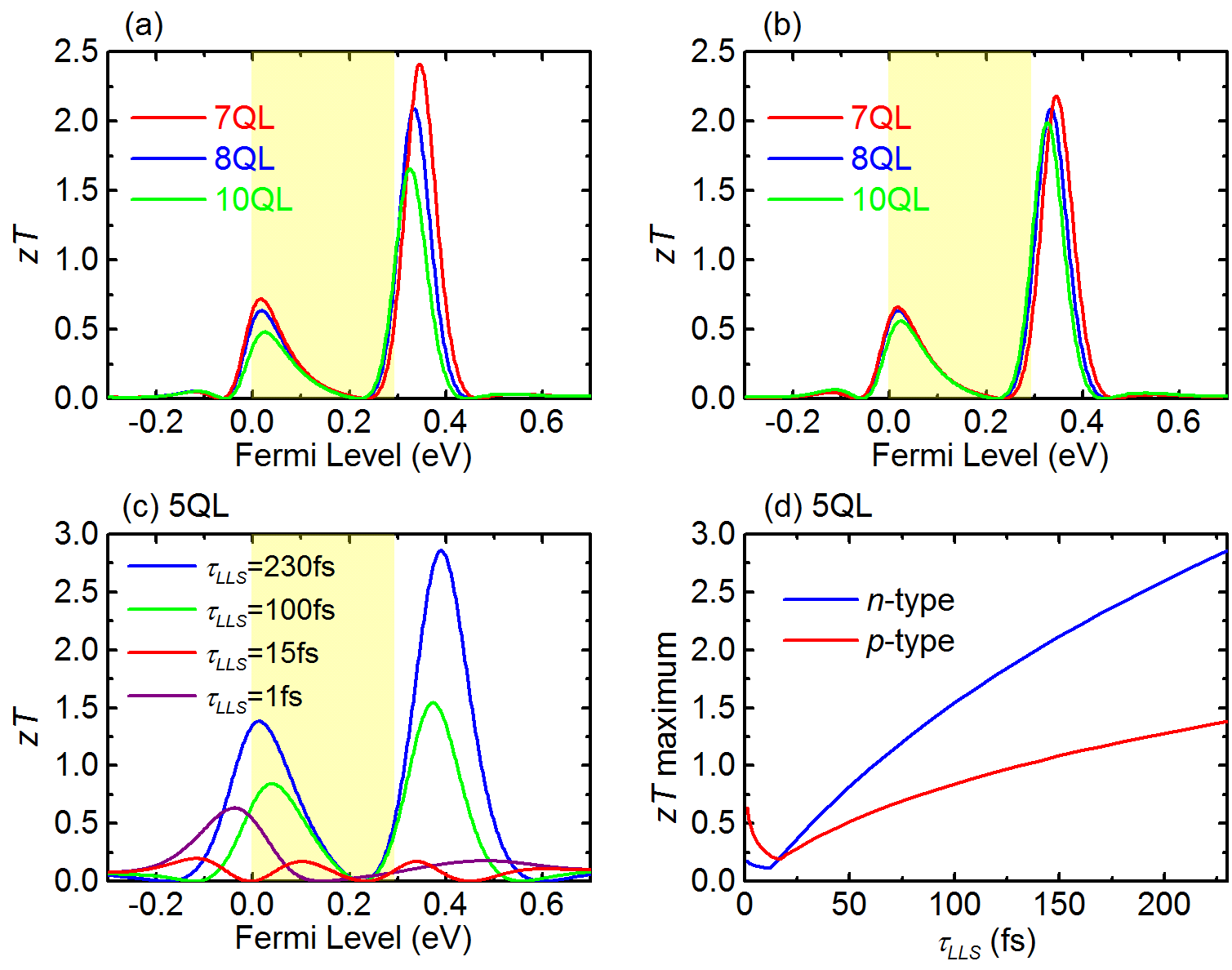}
\caption{\textbf{The thickness dependence of the figure of merit in the Bi}$_{2}$%
\textbf{Se}$_{3}$\textbf{\ thin films.} (\textbf{a}) The room-temperature
figure of merit as a function of Fermi level for the films of different
thickness. (\textbf{b}) The same as (\textbf{a}) except that the lattice
thermal conductance of all the films are set at the value the same as that
of the $8$QL film. (\textbf{c}) The figure of merit of the $5$QL film at
temperature $T=600$ K as a function of the Fermi level.  (\textbf{d}) The $zT
$ maximum of the $5$QL film at temperature $T=600$ K as a function of the
LLS relaxation time for $n$-type ($p$-type) doping, i.e., for $E_{f}>0.2$ eV
($E_{f}<0.2$ eV). The relaxation time of the LLSs in (\textbf{a, b}) is set at $230$ fs and that of the SLSs in (\textbf{a-d}) 1fs.}%
\end{figure}



\newpage

\begin{thebibliography}{99}
\bibitem{Zevenhoven(2011)} Zevenhoven, R. \& Beyene, A. The relative
contribution of waste heat from power plants to global warming. \textit{%
Energy} \textbf{36}, 3754-3762 (2011).

\bibitem{Rowe(1955)} Rowe, D. M. (ed.) CRC \textit{Handbook of
thermoelectrics}. (CRC Press, 1955).

\bibitem{Tritt(2006)} Tritt, T. M. \& Subramanian, M. A. Thermoelectric
materials, phenomena, and applications: A bird's eye view. \textit{Mrs Bull} 
\textbf{31}, 188-194 (2006).

\bibitem{Snyder(2008)} Snyder, G. J. Small thermoelectric generators. 
\textit{Electrochem. Soc. Interface} \textbf{17}, 54 (2008).

\bibitem{Tritt(2011)} Tritt, T. M. Thermoelectric phenomena, materials, and
applications. \textit{Annu. Rev. Mater. Res.} \textbf{41}, 433-448 (2011).

\bibitem{Alam(2013)} Alam, H. \& Ramakrishna, S. A review on the enhancement
of figure of merit from bulk to nano-thermoelectric materials. \textit{Nano
Energy} \textbf{2}, 190-212 (2013).

\bibitem{He(2015)} He, W., Zhang, G., Zhang, X. X., Ji, J., Li, G. Q. \&
Zhao, X. D. Recent development and application of thermoelectric generator
and cooler. \textit{Appl. Energ.} \textbf{143}, 1-25 (2015).

\bibitem{Aswal(2016)} Aswal, D. K., Basu, R. \& Singh, A. Key issues in
development of thermoelectric power generators: High figure-of-merit
materials and their highly conducting interfaces with metallic
interconnects. \textit{Energ. Convers. Manage.} \textbf{114}, 50-67 (2016).

\bibitem{Zhang(2016)} Zhang, Q. H., Huang, X. Y., Bai, S. Q., Shi, X., Uher,
C. \& Chen, L. D. Thermoelectric devices for power generation: Recent
progress and future challenges. \textit{Adv. Eng. Mater.} \textbf{18},
194-213 (2016).

\bibitem{Kane(2005)} Kane, C. L. \& Mele, E. J. Z(2) topological order and
the quantum spin Hall effect. \textit{Phys. Rev. Lett.} \textbf{95}, 146802
(2005).

\bibitem{Bernevig(2006)prl} Bernevig, B. A. \& Zhang, S. C. Quantum spin
hall effect. \textit{Phys. Rev. Lett.} \textbf{96}, 106802 (2006).

\bibitem{Fu(2007)} Fu, L., Kane, C. L. \& Mele, E. J. Topological insulators
in three dimensions. \textit{Phys. Rev. Lett.} \textbf{98}, 106803 (2007).

\bibitem{Moore(2007)} Moore, J. E. \& Balents, L. Topological invariants of
time-reversal-invariant band structures. \textit{Phys. Rev. B} \textbf{75},
121306 (2007).

\bibitem{Qi(2008)} Qi, X. L., Hughes, T. L. \& Zhang, S. C. Topological
field theory of time-reversal invariant insulators. \textit{Phys. Rev. B} 
\textbf{78}, 195424 (2008).

\bibitem{Hasan(2010)} Hasan, M. Z. \& Kane, C. L. Colloquium: Topological
insulators. \textit{Rev. Mod. Phys.} \textbf{82}, 3045-3067 (2010).

\bibitem{Bansil(2016)} Bansil, A., Lin, H. \& Das, T. Colloquium:
Topological band theory. \textit{Rev. Mod. Phys.} \textbf{88} (2016).

\bibitem{Zhang(2009)} Zhang, H. J., Liu, C. X., Qi, X. L., Dai, X., Fang, Z.
\& Zhang, S. C. Topological insulators in Bi$_{2}$Se$_{3}$, Bi$_{2}$Te$_{3}$
and Sb$_{2}$Te$_{3}$ with a single Dirac cone on the surface. \textit{Nat.
Phys.} \textbf{5}, 438-442 (2009).

\bibitem{Chen(2009)} Chen, Y. L., Analytis, J. G., Chu, J. H., Liu, Z. K.,
Mo, S. K., Qi, X. L., Zhang, H. J., Lu, D. H., Dai, X., Fang, Z., Zhang, S.
C., Fisher, I. R., Hussain, Z. \& Shen, Z. X. Experimental realization of a
three-dimensional topological insulator, Bi$_{2}$Te$_{3}$. \textit{Science} 
\textbf{325}, 178-181 (2009).

\bibitem{Xie(2010)} Xie, W. J., He, J., Kang, H. J., Tang, X. F., Zhu, S.,
Laver, M., Wang, S. Y., Copley, J. R. D., Brown, C. M., Zhang, Q. J. \&
Tritt, T. M. Identifying the specific nanostructures responsible for the
high thermoelectric performance of (Bi,Sb)$_{2}$Te$_{3}$ nanocomposites. 
\textit{Nano Lett.} \textbf{10}, 3283-3289 (2010).

\bibitem{Muchler(2013)} Muchler, L., Casper, F., Yan, B. H., Chadov, S. \&
Felser, C. Topological insulators and thermoelectric materials. \textit{%
Phys. Status Solidi-R} \textbf{7}, 91-100 (2013).

\bibitem{Shi(2015)} Shi, H. L., Parker, D., Du, M. H. \& Singh, D. J.
Connecting thermoelectric performance and topological-insulator behavior: Bi$%
_{2}$Te$_{3}$ and Bi$_{2}$Te$_{2}$Se from first principles. \textit{Phys.
Rev. Appl.} \textbf{3} (2015).

\bibitem{ZhangSC(2014)} Xu, Y., Gan, Z. X. \& Zhang, S. C. Enhanced
thermoelectric performance and anomalous Seebeck effects in topological
insulators. \textit{Phys. Rev. Lett.} \textbf{112}, 226801 (2014).

\bibitem{Liang(2016)} Liang, J. H., Cheng, L., Zhang, J., Liu, H. J. \&
Zhang, Z. Y. Maximizing the thermoelectric performance of topological
insulator Bi$_{2}$Te$_{3}$ films in the few-quintuple layer regime. \textit{%
Nanoscale} \textbf{8}, 8863-8870 (2016).

\bibitem{Xia(2009)} Xia, Y., Qian, D., Hsieh, D., Wray, L., Pal, A., Lin,
H., Bansil, A., Grauer, D., Hor, Y. S., Cava, R. J. \& Hasan, M. Z.
Observation of a large-gap topological-insulator class with a single Dirac
cone on the surface. \textit{Nat. Phys.} \textbf{5}, 398-402 (2009).

\bibitem{Hsieh(2009)} Hsieh, D., Xia, Y., Qian, D., Wray, L., Dil, J. H.,
Meier, F., Osterwalder, J., Patthey, L., Checkelsky, J. G., Ong, N. P.,
Fedorov, A. V., Lin, H., Bansil, A., Grauer, D., Hor, Y. S., Cava, R. J. \&
Hasan, M. Z. A tunable topological insulator in the spin helical Dirac
transport regime. \textit{Nature} \textbf{460}, 1101-1105 (2009).

\bibitem{Sun(2012)} Sun, Y. F., Cheng, H., Gao, S., Liu, Q. H., Sun, Z. H.,
Xiao, C., Wu, C. Z., Wei, S. P. \& Xie, Y. Atomically thick bismuth selenide
freestanding single layers achieving enhanced thermoelectric energy
harvesting. \textit{J. Am. Chem. Soc.} \textbf{134}, 20294-20297 (2012).

\bibitem{Kresse(1996)a} Kresse, G. \& Furthmuller, J. Efficiency of
ab-initio total energy calculations for metals and semiconductors using a
plane-wave basis set. \textit{Comp. Mater. Sci.} \textbf{6}, 15-50 (1996).

\bibitem{Kresse(1995)} Kresse, G. \textit{Ab initio} molecular dynamics for
liquid-metals. \textit{J. Non-Cryst Solids} \textbf{193}, 222-229 (1995).

\bibitem{Kresse(1996)b} Kresse, G. \& Furthmuller, J. Efficient iterative
schemes for \textit{ab initio} total-energy calculations using a plane-wave
basis set. \textit{Phys. Rev. B} \textbf{54}, 11169-11186 (1996).

\bibitem{Perdew(1992)a} Perdew, J. P., Chevary, J. A., Vosko, S. H.,
Jackson, K. A., Pederson, M. R., Singh, D. J. \& Fiolhais, C. Atoms,
molecules, solids, and surfaces: Applications of the generalized gradient
approximation for exchange and correlation. \textit{Phys. Rev. B} \textbf{46}%
, 6671-6687 (1992).

\bibitem{Perdew(1992)b} Perdew, J. P. \& Wang, Y. Pair-Distribution Function
and Its Coupling-Constant Average for the Spin-Polarized Electron-Gas. 
\textit{Phys. Rev. B} \textbf{46}, 12947-12954 (1992).

\bibitem{Grimme(2010)} Grimme, S., Antony, J., Ehrlich, S. \& Krieg, H. A
consistent and accurate ab initio parametrization of density functional
dispersion correction (DFT-D) for the 94 elements H-Pu. \textit{J. Chem.
Phys.} \textbf{132} (2010).

\bibitem{Grimme(2011)} Grimme, S., Ehrlich, S. \& Goerigk, L. Effect of the
damping function in dispersion corrected density functional theory. \textit{%
J. Comput. Chem.} \textbf{32}, 1456-1465 (2011).

\bibitem{Mahan(1996)} Mahan, G. D. \& Sofo, J. O. The best thermoelectric. 
\textit{Proc. Natl. Acad. Sci. USA} \textbf{93}, 7436-7439 (1996).

\bibitem{Callaway(1959)} Callaway, J. Model for lattice thermal conductivity
at low temperatures. \textit{Phys. Rev.} \textbf{113}, 1046-1051 (1959).

\bibitem{Navratil(2004)} Navratil, J., Horak, J., Plechacek, T., Kamba, S.,
Lost'ak, P., Dyck, J. S., Chen, W. \& Uher, C. Conduction band splitting and
transport properties of Bi$_{2}$Se$_{3}$. \textit{J. Solid State Chem.} \textbf{177},
1704-1712 (2004).

\bibitem{Taskin(2012)} Taskin, A. A., Sasaki, S., Segawa, K. \& Ando, Y.
Manifestation of topological protection in transport properties of epitaxial
Bi$_{2}$Se$_{3}$ thin films. \textit{Phys. Rev. Lett.} \textbf{109}, 066803 (2012).

\bibitem{Luo(2012)} Luo, X., Sullivan, M. B. \& Quek, S. Y. First-principles
investigations of the atomic, electronic, and thermoelectric properties of
equilibrium and strained Bi$_{2}$Se$_{3}$ and Bi$_{2}$Te$_{3}$ including van
der Waals interactions. \textit{Phys. Rev. B} 86, 184111 (2012).

\bibitem{Goldsmid(book)} Goldsmid, H. J. \textit{Introduction to
thermoelectricity}. Vol. 121 (Springer, 2009).

\bibitem{Mermin(book)} Aschcroft, N. W. \& Mermin, N. D. \textit{Solid state
physics}. (Holt-Saunders, 1976).
\pagebreak
\end{thebibliography}
\end{document}


\title{Supplementary Information of Enhanced Thermoelectric Performance in Thin Films of
Three-Dimensional Topological Insulators}
\author{T. H. Wang and H. T. Jeng}
\affiliation{Department of Physics, National Tsing Hua University, 101 Section 2 Kuang Fu
Road, Hsinchu,Taiwan 30013,R.O.C.}

\begin{abstract}
\bigskip \bigskip \bigskip \bigskip
\noindent\textbf{Table of contents: }
\bigskip 

\noindent%
\begin{tabular}{ll}
I. & \textbf{Estimation of the LLS relaxation time} \\ 
II. & \textbf{TE performance in the limit of large LLS relaxation time} \\
II. & \textbf{Further Enhancement by shortening the SLS relaxation time at room temperature}%
\end{tabular}
\end{abstract}

\maketitle

\newpage
\section{Estimation of the LLS relaxation time}
\begin{figure}[H]
\includegraphics [width=17cm]{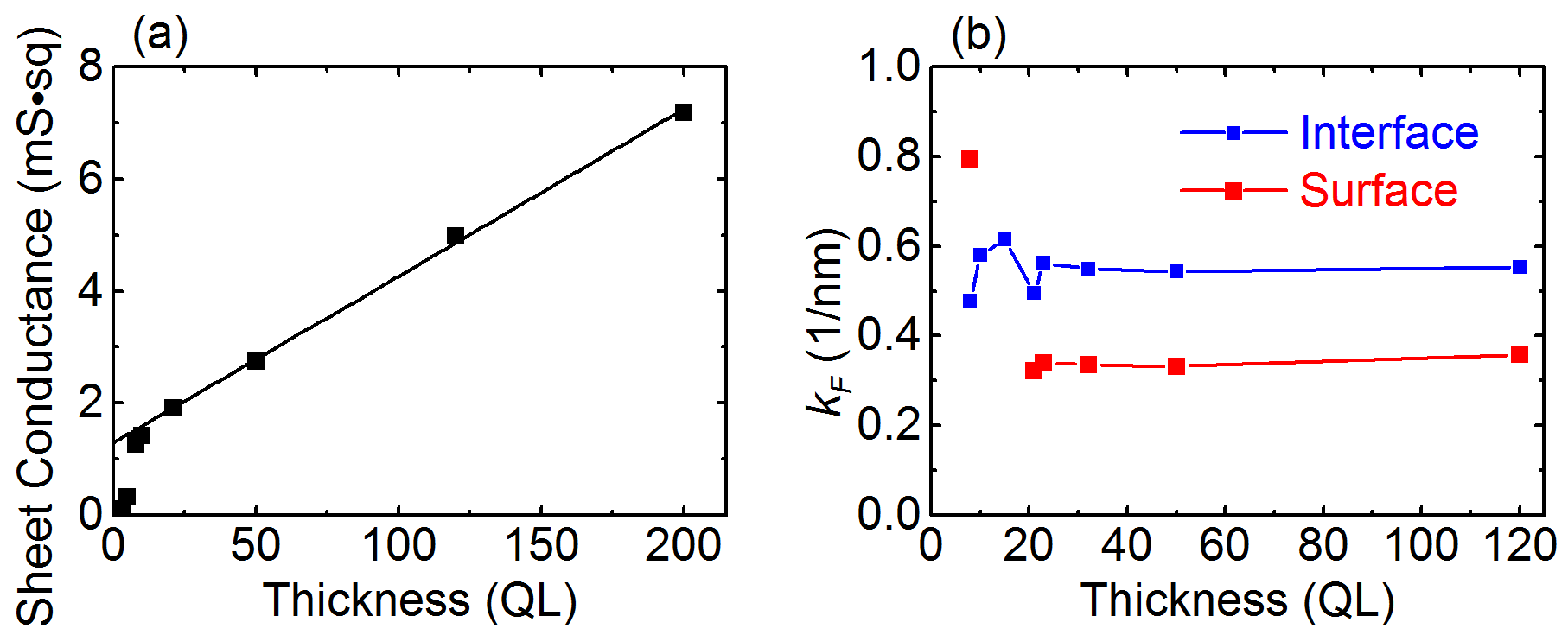}
\caption{\textbf{The room-temperature sheet conductance and Fermi wave number for the films of different thickness measured in ref. \citenum{Taskin(2012)}}. \textbf{(a)} The sheet conductance. The dots are the experimental data points and the straight line is a linear fit to the data points of thickness larger than $20$ QLs. \textbf{(b)} The surface (interface) Fermi wave number obtained from the Fourier spectra of the Shubnikov-de Haas oscillations.}%
\end{figure}

$\quad $
The relaxation time of the LLSs (i.e., the TSSs in the bulk gap) of the Bi$%
_{2}$Se$_{3}$ films can be estimated by comparing the measured sheet
conductance in ref. \citenum{Taskin(2012)} with our calculated results. Figure S1a shows the
measured sheet conductance of the samples with different thickness at room
temperature. As can be seen, the sheet conductance increases linearly with
the film thickness for the samples of thickness $d\geq 8$ QLs. The linear
dependence of the conductance on the film thickness implies that the carrier
concentration and hence the Fermi energy of theses samples in the region
away from the surface are nearly the same. The latter, according to the Mott
criterion, should be near the CBE\cite{Brahlek(2015)}. The intercept at zero
thickness, $\sigma (0)$, should be be nearly equal to the sheet conductance
of the LLSs. The conductance drops dramatically when the film thickness
decreases from $8$ QLs to $5$ QLs indicating the dramatic decrease of the
LLS relaxation time.\ This is an experimental evidence that the scattering
between the TSSs on the opposite sides of the thin film becomes effective
when the film thickness decreases from $8$ QLs to $5$ QLs. It is worth
noting that the large-angle scattering between the TSSs on the same side is
suppressed while that between the TSSs on the opposite sides is enhanced
because the spin polarizations of the TSSs at different sides circulate the
circular energy contour oppositely. The linear dependence of the conductance
on the film thickness also indicates that the LLS conductances and hence the
surface (interface) Fermi energies of the samples whose thickness is not
smaller than $8$ QLs should be almost the same. The surface (interface)
Fermi wave number $k_{F}$ can be obtained from the Fourier transform of the
Shubnikov-de Haas (SdH) oscillations reported in ref. \citenum{Taskin(2012)}. In general, there are two oscillation frequencies
corresponding to the Fermi wave numbers at different sides of the Bi$_{2}$Se$%
_{3}$ film (i.e., the surface and the interface between the Bi$_{2}$Se$_{3}$
film and the substrate) as plotted in Fig. S1b. As can be seen, the SdH
frequency corresponding to $k_{F}\sim 0.55$ ($1/$nm) always exists while the
other one does not. Therefore, the former should be resulted from the
interface while the latter the surface. The surface time-dependent downward
band bending of the Bi$_{2}$Se$_{3}$ has been commonly observed. For
example, it has been seen, when the Bi$_{2}$Se$_{3}$ surface is exposed to
the carbon monoixide\cite{Bianchi(2011)}, the water vapor\cite{Benia(2011)},
the ambient environment\cite{Bianchi(2010)} and even the residual gas in
vacuum\cite{King(2011)}. It can be the surface downward band bending effect
so that the SdH oscillation of the surface TSS is not observed for the
samples of thickness $10$ QLs and $15$ QLs\cite{Brahlek(2015)}, and the
magnitude of the surface Fermi wave number of the $8$QL samples is much
larger than the samples of thickness larger than $20$ QLs. When the surface
Fermi level is higher than the CBE due to the surface downward band bending,
the conductance contributed from the TSSs near the surface should be much
smaller than that near the interface\cite{Brahlek(2015)}. In the conductance
measurement in ref. ~\citenum{Taskin(2012)}, the conductance contributed from the TSS near the
surface should be negligible, otherwise the total conductance shown in Fig.
S1a will not exhibit a linear dependence on the film thickness. Using the
Fermi wave number $k_{F}=0.55$ ($1/$nm), we get the Fermi level $E_{f}=145$
meV. Comparing the LLS conductance $\sigma (0)=1.29$ mS$\cdot $sq, i.e. the
intercept of the straight line in Fig. S1a, with our calculation of the LLS
conductance $\sigma _{LLS}$, we conclude that the LLS relaxation time is
about $230$ fs.

\section{TE performance in the limit of large LLS relaxation time}
\begin{figure}[H]
\includegraphics [width=8.5cm]{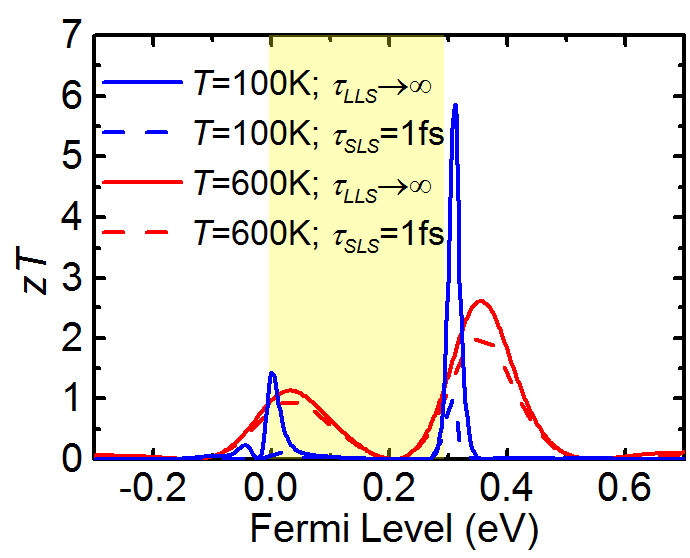}
\caption{\textbf{The figure of merit of the} $8$\textbf{QL Bi}$_{2}$\textbf{Se}$_{3}$ \textbf{as a function of the Fermi level at
temperatures} $T=100$ \textbf{K (blue lines) and} $T=600$ \textbf{K (red lines).} The ratio of
the LLS to the SLS relaxation time $\tau _{LLS}/\tau _{SLS}$ is set to be $%
230$. The solid lines are the results in the limit of infinite LLS
relaxation time and the dashed lines are for the case of $\tau _{SLS}=1$ fs.}%
\end{figure}
$\quad $
From Figs. 3a (3k) and 4a (4k) in the main text, we find the peak value of
the figure of merit at low temperature ($T=100$ K) is much smaller than that
at high temperature ($T=600$ K). However, thus far, we set the LLS
relaxation time $\tau _{LLS}=230$ fs, which, to some extent, is determined
by the sample quality. The LLS relaxation time could be further extended by
improving the surface quality (or more precisely, the sample quality in the
region $3$ QLs near the surface according to our calculation of the TSS wave
functions). If the LLS relaxation time is so long that the electronic
thermal conductance is much larger than the lattice thermal conductance, the
peak value of the figure of merit at low temperature could be higher than
that at high temperature. In the limit of infinite LLS relaxation time, the
ratio $\sigma /\kappa $, Seebeck ceofficeint $S$, and hence the figure of
merit $zT$ do not correlate with $\tau _{LLS}$ or $\tau _{SLS}$ for a given
relaxation time ratio $r_{\tau }\equiv \tau _{LLS}/\tau _{SLS}$. Figure S2
shows the resultant figure of merit, of which the $r_{\tau }$ is set to be $%
230$. As can be seen, the maximum of the figure of merit at $100$ K is more
than twice larger than that at $600$ K in the limit of infinite LLS
relaxation time (Solid lines). As a comparison, the dashed lines shows the
case of $\tau _{SLS}=1$ fs, and the maximum of the figure of merit at $100$
K is more than twice smaller than that at $600$ K.

\section{Further Enhancement by shortening the SLS relaxation time at room temperature}
\begin{figure}[H]
\includegraphics [width=17.78cm]{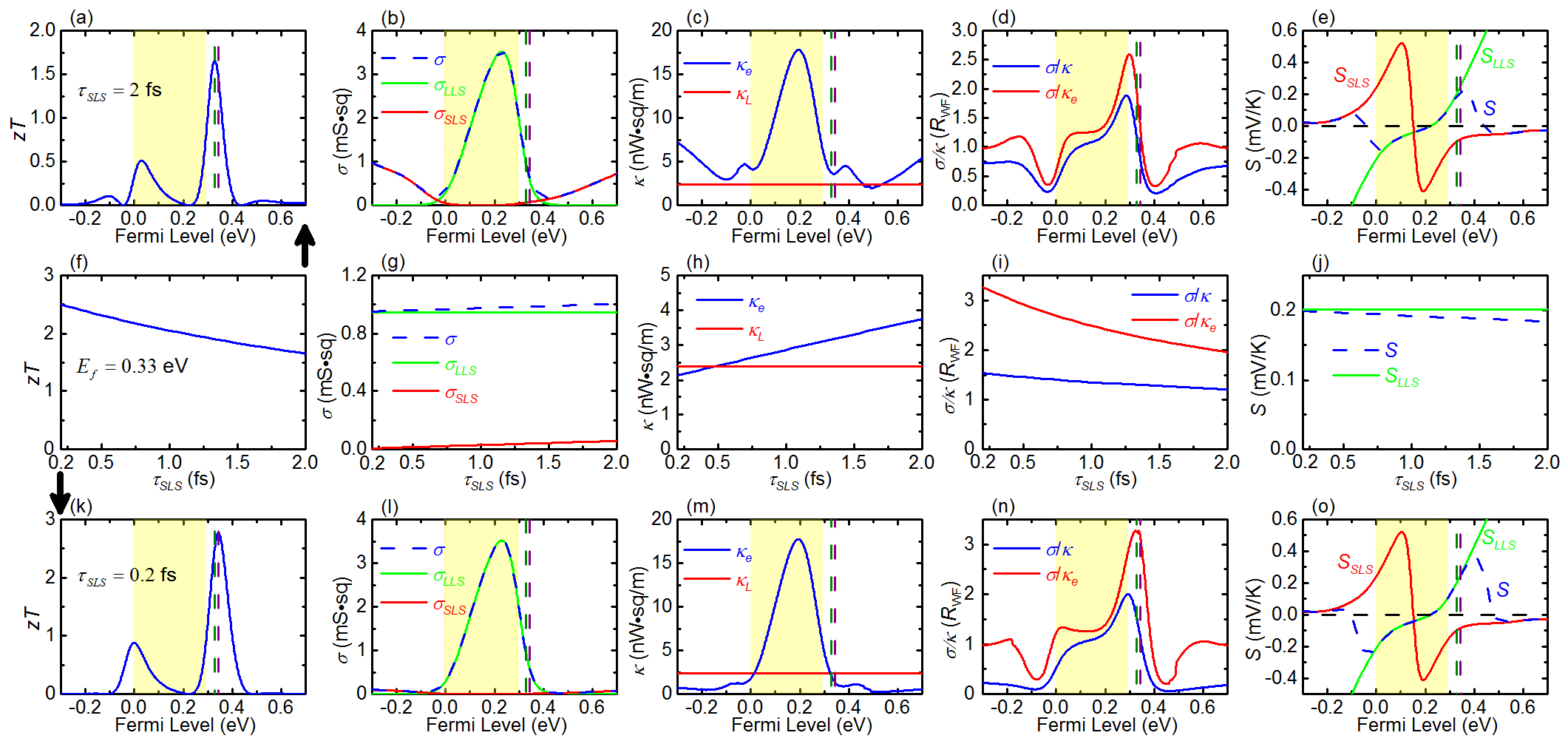}
\caption{\textbf{The thermoelectric parameters of the }$8$\textbf{QL Bi}$_{2}$\textbf{Se}$_{3}$\textbf{\ at temperature }$T=300$\textbf{\ K.} (\textbf{a-e}) The thermoelectric parameters as functions of the Fermi level for the SLS relaxation time being set at $\tau_{SLS}=2$ fs and (\textbf{k-o}) those for $\tau_{SLS}=0.2$ fs. In (\textbf{a-e}) and (\textbf{k-o}), the vertical green line and the vertical purple line indicate, respectively, the P3 energy for $\tau_{SLS}=2$ fs and that for $\tau_{SLS}=0.2$ fs. (\textbf{f-j}) The thermoelectric parameters as functions of SLS relaxation time when the Fermi level is set at $0.33$ eV, the P3 energy for $\tau_{SLS}=2$ fs.}

\end{figure}